\documentclass[a4paper, twocolumn, 11pt, accepted=2022-04-05]{quantumarticle}
\pdfoutput=1
\usepackage[utf8]{inputenc}
\usepackage[english]{babel}
\usepackage[T1]{fontenc}
\usepackage{amsmath}
\usepackage{hyperref}
\usepackage{cite}

\usepackage{tikz}
\usepackage{lipsum}

\begin{document}

\title{Flow of time during energy measurements and the resulting time-energy uncertainty relations}

\author{I. L. Paiva}
\affiliation{Faculty of Engineering and the Institute of Nanotechnology and Advanced Materials, Bar-Ilan University, Ramat Gan 5290002, Israel}
\orcid{0000-0002-0416-3582}

\author{A. C. Lobo}
\affiliation{Institute of Exact and Biological Sciences, Federal University of Ouro Preto, Ouro Preto, Minas Gerais 35400-000, Brazil}

\author{E. Cohen}
\affiliation{Faculty of Engineering and the Institute of Nanotechnology and Advanced Materials, Bar-Ilan University, Ramat Gan 5290002, Israel}
\orcid{0000-0001-6198-0725}

\begin{abstract}
Uncertainty relations play a crucial role in quantum mechanics. Well-defined methods exist for the derivation of such uncertainties for pairs of observables. Other approaches also allow the formulation of time-energy uncertainty relations, even though time is not an operator in standard quantum mechanics. However, in these cases, different approaches are associated with different meanings and interpretations for these relations. The one of interest here revolves around the idea of whether quantum mechanics inherently imposes a fundamental minimum duration for energy measurements with a certain precision. In our study, we investigate within the Page and Wootters timeless framework how energy measurements modify the relative ``flow of time'' between internal and external clocks. This provides a unified framework for discussing the subject, allowing us to recover previous results and derive new ones. In particular, we show that the duration of an energy measurement carried out by an external system cannot be performed arbitrarily fast from the perspective of the internal clock. Moreover, we show that during any energy measurement the evolution given by the internal clock is non-unitary.
\end{abstract}

\maketitle

\section{Introduction}

In standard quantum mechanics, while the canonical conjugate variables of position and momentum are treated as observables, the same does not hold for the pair time-energy. While energy (associated with a Hamiltonian) is an observable, time is treated as an external parameter. Since time and position are part of the same object in relativistic theories, the lack of symmetry in their status in quantum mechanics seems to make the incompatibility of this theory with relativity deeper than the incompatibility of earlier non-relativistic theories. An answer to this conundrum was the development of the relativistic quantum field theory, which is, to this day, the main tool to describe the standard model of elementary particles. This approach ``downgrades'' the status of the position observable to a parameter, putting it alongside time, while ``upgrading'' quantum particles to the status of fields. In turn, this gives place to the arena of the ``classical'' Minkowski spacetime, upon where the quantum relativistic fields are defined. Yet, this move does not solve many fundamental issues present in quantum mechanics. In particular, the problem of constructing a consistent quantum theory of gravity remains.

A more recent venue to reconsider this problem is to look back at Einstein's operational approach that led to the discovery of special relativity in the first place. He achieved a better understanding of space and time not by asking deep metaphysical questions, such as what space and time are, but by answering much more seemingly mundane questions, like how physicists in relative motion synchronize the measurements of length and time between events with rods and clocks. An approach to quantum mechanics that abides by this operational perspective has been coined \textit{relational quantum mechanics}. It may become instrumental in describing spacetime and gravity quantum mechanically \cite{rovelli1996relational}. The main idea is to recognize that clocks and rulers are physical objects and, as such, they must obey quantum mechanical laws. Also, physical quantities are always given relative to some reference object or quantity, and so one is led to the concept of \textit{quantum reference frames} \cite{aharonov1984quantum, bartlett2007reference, angelo2011physics}. As it will be seen, this approach is central to the discussion here.

A related question to the aforementioned relativistic asymmetry of quantum mechanics concerns the existence of a time-energy uncertainty relation. On the one hand, this type of relation is expected to exist in view of relativistic ideas together with Heisenberg's uncertainty principle, which establishes an uncertainty relation for position and momentum. However, standard approaches for obtaining uncertainty relations, which are appropriate for the study of the relation between two observables \cite{heisenberg1985anschaulichen, kennard1927quantenmechanik, robertson1929uncertainty, schrodinger1930zum, maccone2014stronger}, do not work for time-energy. As a result, there exist multiple ways to address this problem \cite{bauer1978time, busch1990energy, busch1990energy2, dodonov2015energy}. For instance, Mandelstam and Tamm introduced a time-energy uncertainty relation based on the fact that, in many cases, for a general observable $O$, the ratio $\Delta O/\langle \dot{O} \rangle$, where $\Delta O$ is the variance of $O$ and $\dot{O}$ is its time derivative, defines an interval of time \cite{mandelstam1945uncertainty}. Also, Anandan and Aharonov presented a geometric time-energy uncertainty relation based on the concept of orthogonal-time, defined as the time that it takes for the state of a system to become orthogonal to its initial state \cite{anandan1990geometry}. A similar idea to the latter was later examined by Margolus and Levitin in Ref. \cite{margolus1998maximum}.

However, the approach that is of particular interest here and whose first studies precede the aforementioned two has to do with time-energy relations associated with the process of measurements of energy. More specifically, relations that determine how fast such measurements can be made given a certain desirable precision. First, Landau and Peierls argued in favor of an uncertainty relation with a particular model for measuring energy \cite{landau1931erweiterung}. However, Aharonov and Bohm showed that their model did not provide an optimal energy measurement scheme and, in fact, it is possible to design energy measurements that can be performed arbitrarily fast \cite{aharonov1961time}. Nevertheless, a vast number of measurements are bounded by an uncertainty relation. Investigating the reasons behind it, Aharonov, Massar, and Popescu showed that arbitrarily fast measurements require prior knowledge about the Hamiltonian \cite{aharonov2002measuring}. In case the Hamiltonian is unknown, as they showed, some time must be spent in the estimation of it. Only after that, the measurement of energy can be performed arbitrarily fast.

In common, these discussions considered measurements carried out by external systems to the system of interest. However, as analyzed by Aharonov and Reznik, one can also consider the case where the system's energy is measured by a part of it \cite{aharonov2000weighing}. They suggested that, in this case, the measurement can never be performed arbitrarily fast. However, Massar and Popescu showed that if one considers the proper time (i.e., internal time) of the system of interest, there exists no limit on the speed of the measurement \cite{massar2005measurement}.

These studies, directly or indirectly, involve the idea of \textit{quantum frames of reference} and, in particular, clock frames. The point in common is that they all use a physical system to represent a clock. However, the way these approaches were constructed makes it non-trivial to discuss the different scenarios (i.e., internal or external measurements with an analysis of the internal or external time) in a unified way. In fact, each of these results relies on specific measurement schemes. Their approach included \textit{a single clock system}. Because of that, in the works that studied both internal and external times (Refs. \cite{aharonov2000weighing, massar2005measurement}), an analysis of the composed state of the relevant systems after the end of the von Neumann interaction was carried out to draw conclusions about the duration of the measurement according to the two distinct notions of time.

Regarding the idea of quantum clocks, Page and Wootters introduced a framework for the study of timeless quantum mechanics \cite{page1983evolution}. Assuming the state of the joint system composed by a clock and a system of interest satisfy a certain constraint, known as the Wheeler-DeWitt equation, they showed that the unitary Schr\"odinger dynamics of the system of interest could be recovered in their timeless framework. This approach was further developed in Refs. \cite{wootters1984time, giovannetti2015quantum, marletto2017evolution, smith2019quantizing, hohn2019trinity, giacomini2019quantum, diaz2019history, diaz2019history2, hohn2020switch, castro2020quantum, smith2020quantum, ballesteros2021group, trassinelli2021conditional, paiva2021nonlocality, baumann2021page} and is similar to other approaches in the literature, e.g., in Refs. \cite{reisenberger2002spacetime, hellmann2007multiple}. It has also been realized experimentally \cite{moreva2014time, moreva2017quantum}.

Among the studies in this area, multiple clocks and how quantum systems evolve from their perspective were considered \cite{hohn2019trinity, hohn2020switch, castro2020quantum}. Inspired by it, we introduce the discussion about the passage of time during a von Neumann measurement (or pre-measurement since the final ``collapse'' is not involved) of energy in the timeless framework. This allows us to analyze how the ``flow of time'' changes during the measurement in an internal or external clock and from the perspective of each of them. Besides the interesting aspects of this analysis on its own, it also provides a unified and simplified framework for the study and understanding of time-energy uncertainty relations associated with the process of energy measurements, as formulated by Landau and Peierls \cite{landau1931erweiterung} and further developed in Refs. \cite{aharonov1961time, aharonov2002measuring, massar2005measurement}.

One of the main differences proposed by our approach is the explicit use of \textit{two physical clocks}: one internal and the other external to the system of interest. As a result, we do not need to rely on analysis of phase factors multiplying the wave function of the system after the measurement is completed, as done in previous works. In addition, as already mentioned, the model we consider allows us to observe how the measurement interaction changes the relative flow of time between the different clocks \textit{during} the measurement. Furthermore, with this unified framework, we can also draw conclusions about the existence or absence of a time-energy uncertainty relation for the internal time when the measurement is carried out by an external system and for the external time when the measurement is carried out by an internal subpart of the system of interest. To the best of our knowledge, the existence or not of a time-energy uncertainty relation in these two scenarios was not known prior to the present work.

Another feature revealed by our results is the appearance of non-unitarity from the internal's clock perspective regardless of the measurement being conducted by an external or an internal system. This is another contribution of this article. In previous discussions of time-energy uncertainty relations, the particularities of the models and the way they were solved did not allow for this property to be observed. Furthermore, in the context of the Page and Wootters framework, to the best of our knowledge, this work is the first to point out that the effective Hamiltonian of a system from a clock's perspective is not always Hermitian.

The article is organized as follows. In Section \ref{sec:pw}, we present an overview of Page and Wootters framework. Following that, in Section \ref{sec:measurement}, we introduce our model for the measurement of energy carried out by both an external system or an internal part of the system of interest. This allows to analyze the relative flow of time between the different clocks. This leads to a two-fold study: the subsequent time-energy uncertainty relations in Section \ref{sec:uncertainty} and the emergence of non-unitarity in Section \ref{sec:nonunitarity}. Finally, we discuss the results presented here as well as some remaining questions that deserve further examination in Section \ref{sec:conclusion}.

\section{Quantum mechanics in the timeless framework}
\label{sec:pw}

The timeless framework consists of a clock system, whose state is given by a vector in a Hilbert space $\mathcal{H}_A$, and the rest of the system, represented by a state in a Hilbert space $\mathcal{H}_R$, whose evolution is studied. The joint system $|\Psi\rangle\rangle \in \mathcal{H}_A\otimes\mathcal{H}_R$ is assumed to be closed and, hence,
\begin{equation}
    H_T|\Psi\rangle\rangle = 0,
    \label{eq:constraint}
\end{equation}
where $H_T$ is the total Hamiltonian acting on systems $A$ and $R$. This equation is known as the Wheeler-DeWitt equation. Observe that the imposition of $|\Psi\rangle\rangle$ being an eigenstate with null eigenvalue by Eq. \eqref{eq:constraint} is not as restrictive as it seems \cite{giovannetti2015quantum}. In fact, Hamiltonians that differ by constant terms are physically equivalent, which is associated with quantum states being defined up to a global phase. Then, Eq. \eqref{eq:constraint} just reads as $|\Psi\rangle\rangle$ not evolving with respect to an external time.

The clock system should have an observable $T_A$ associated with its time. It is desirable that such operator is covariant under translations generated by the Hamiltonian $H_A$, i.e., $|t_0+t_A\rangle = e^{-i H_A t_A/\hbar} |t_0\rangle$, where $|t_0\rangle$ and $|t_0+t_A\rangle$ are taken to be clock states. This does not require $T_A$ to be a self-adjoint canonical conjugate to $H_A$. In fact, it is possible to construct $T_A$ as a positive operator-valued measure (POVM), which means that the clock states are not necessarily eigenstates of $T_A$ \cite{busch199eoperational, busch2016quantum, loveridge2019relative}. Such operators obtained from these constructions are symmetric but need not be self-adjoint. Here, however, for simplicity we assume an \textit{ideal clock}, i.e., $[T_A,H_A]=i\hbar I$, $T_A|t_A\rangle = t_A|t_A\rangle$, and $H_A=-i\hbar \partial/\partial t_A$. Although the Hamiltonians of these clocks are unbounded from bellow and, hence, unrealistic, they help us avoid technicalities associated with real clocks \cite{salecker1958quantum, peres1980measurement} while providing approximations of them \cite{hartle1988quantum, singh2018modeling}. Thus, corrections to the results presented here are expected when dealing with non-ideal clocks.

Moreover, let $H_R$ denote the Hamiltonian of the system of interest and $H_{int}(T_A)$ represent the time dependent term of the evolution of system $R$ set by clock $A$, which is an interaction between $A$ and $R$, we have
\begin{equation}
    H_T = H_A + H_R + H_{int}(T_A).
    \label{eq:pw-h}
\end{equation}
Observe that, in particular, $H_{int}$ is being assumed to be independent of $H_A$. Replacing Eq. \eqref{eq:pw-h} in Eq. \eqref{eq:constraint}, applying a scalar product by an eigenstate $|t_A\rangle$ of $T_A$ on the left, and defining $|\psi(t_A)\rangle \equiv \langle t_A|\Psi\rangle\rangle$, it holds that
\begin{equation}
    i\hbar \frac{\partial}{\partial t_A} |\psi(t_A)\rangle = \left[H_R + H_{int}(t_A)\right]|\psi(t_A)\rangle,
    \label{eq:schrod-general}
\end{equation}
which is the Schr\"odinger equation that gives the evolution of system $R$ with respect to the time measured by clock $A$. Then, the usual unitary evolution of a quantum system is recovered from the static picture introduced by Page and Wootters.

As a result, $|\Psi\rangle\rangle$ can be written as
\begin{equation}
    |\Psi\rangle\rangle = \int dt_A \, |t_A\rangle \otimes |\psi(t_A)\rangle.
\end{equation}
Because $|\Psi\rangle\rangle$ contains information about $|\psi(t_A)\rangle$ at every $t_A$, it is referred to as the \textit{history state}.

The results in Refs. \cite{hohn2019trinity, hohn2020switch, castro2020quantum} are central to this work. In their approach, system $R$ is assumed to contain a clock $B$, and the rest of it is simply referred to as system $S$. Then, while system $B$ gives the internal time of the system $R=B+S$, system $A$ provides time as observed by an external system. In this case, the total Hamiltonian is
\begin{equation}
    H_T = H_A + H_B + H_S + H_{int}(T_B),
\end{equation}
where $T_B$ is the time operator associated with clock $B$. Although the time dependent term $H_{int}$ can be taken to be a function of both $T_A$ and $T_B$ in a more general scenario, we assume it does not depend on $T_A$ for simplicity. However, nothing significant changes in our analysis if $H_{int}$ is also a function of $T_A$.

With the previous $H_T$, the analog of the Schr\"odinger equation becomes
\begin{equation}
    i\hbar \frac{\partial}{\partial t_A} |\psi(t_A)\rangle = \left(H_B + H_S + H_{int}(T_B)\right) |\psi(t_A)\rangle.
    \label{eq:general-2clocks}
\end{equation}
This implies that the effective Hamiltonian acting on system $R$ is
\begin{equation}
    H_{eff}^A \equiv H_B + H_S + H_{int}(T_B).
    \label{eq:ef-hamiltonian}
\end{equation}
Even though this result is a particular case of Eq. \eqref{eq:schrod-general}, it is of special interest here because it allows us to study the relation between the ``flow of time'' in clocks $A$ and $B$.

While the designation of which clock is internal or external to the system of interest seems to be arbitrary up until now, they will acquire a more concrete meaning in the next section, where measurements of energy are studied. Furthermore, in the next section, the evolution will be considered not only from clock $A$'s perspective but also in the frame of clock $B$.

\section{Measurements of energy with quantum clocks}
\label{sec:measurement}

\subsection{Measurement carried out by an external system}

\begin{figure}
    \centering
    \includegraphics[width=\columnwidth]{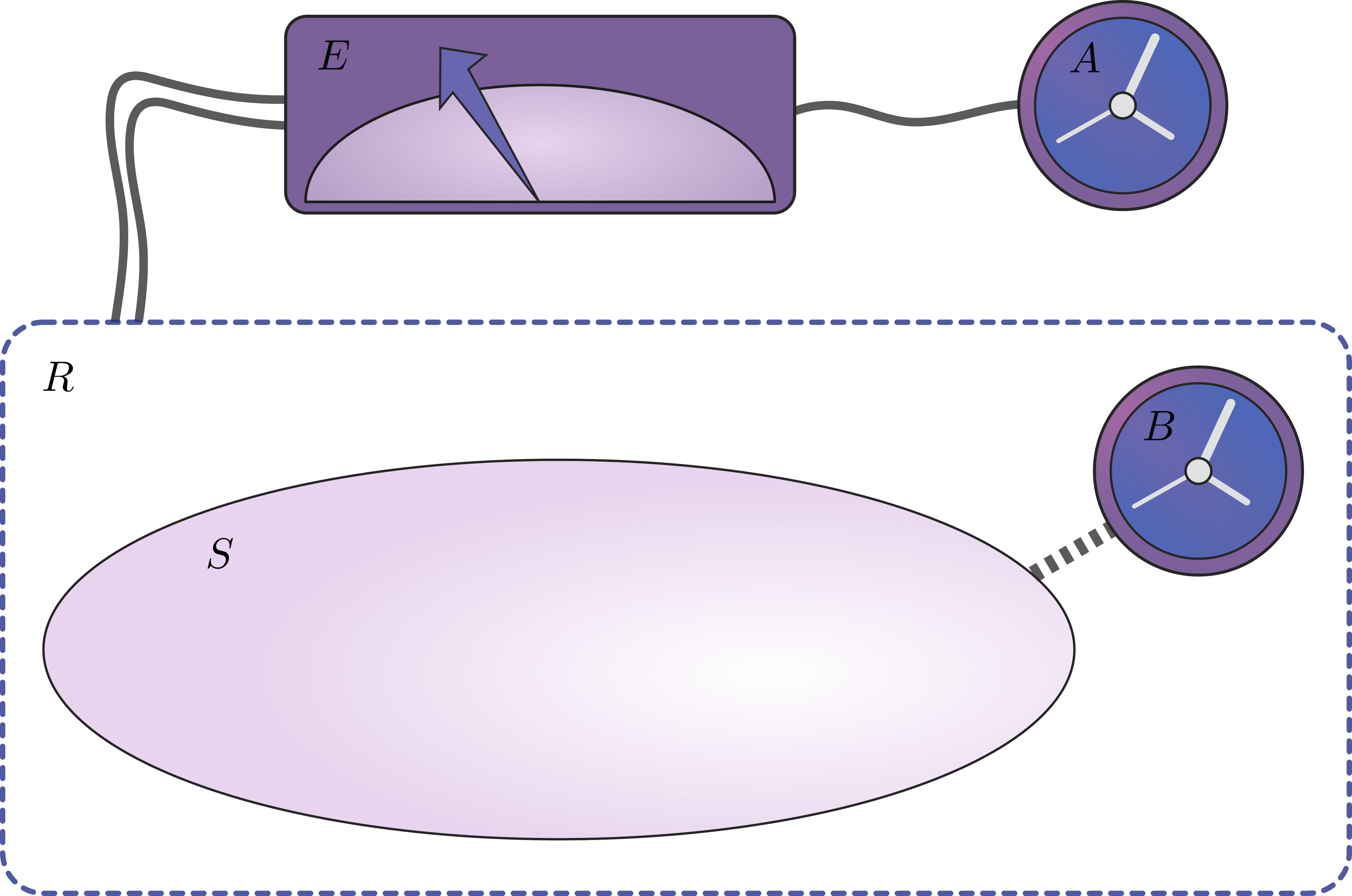}
    \caption{Representation of the relevant parts in a measurement of energy carried out by an external system. The external apparatus $E$ uses an external clock $A$ for time reference during the measurement of energy of system $R$, composed of an internal clock $B$ and system $S$.}
    \label{fig:ext-measurement}
\end{figure}

In this subsection, we study the von Neumann measurement of the total energy of system $R$ in the Page and Wootters timeless framework. Specifically, we consider the case where the measurement is carried out by an external system. Then, in addition to the systems already introduced up until now, we also consider an external pointer system $E$, and the history state $|\Psi\rangle\rangle$ is an element of the space $\mathcal{H}_A \otimes \mathcal{H}_R \otimes \mathcal{H}_E$. This scenario is illustrated in Fig. \ref{fig:ext-measurement}. For simplicity, the free evolution of system $E$ will be neglected.

The measurement interaction can be represented by the von Neumann interaction $H_{VN} \equiv g(T_A) H_R P_E$, where $H_R\equiv H_B+H_S+H_{int}(T_B)$, $P_E$ is the conjugate momentum of the pointer $E$, and $g$ is a non-negative function that differs from zero exclusively during the duration of the measurement. For notation purposes, we assume the measurement starts in clock $A$ at $t_A=0$ and ends at $t_A=\tau$. Moreover, we let
\begin{equation}
    \int_{-\infty}^{\infty} g(t) dt = \int_0^\tau g(t) dt = K,
\end{equation}
where $K$ is a positive real constant associated with the strength (hence, the precision) of the measurement.

In addition, we assume that $P_E$ only takes non-negative values. As we shall see, the duration of the measurement in clock $B$ will linearly depend on $P_E$. Thus, this condition will assure that the clocks considered here are ``good'' clocks in the sense that they do not move backward in time and, in particular, that the duration of the measurement is non-negative. While this requirement on $P_E$ means that the measurement device's pointer cannot be well-localized in an ideal way, it does not forbid it to be localized up to some order. In fact, if $P_E$ has a sufficient large average value, it is possible for the measurement device's pointer to have a small variance since its state can be built in such a way that $P_E$'s variance is large.

With that, the total Hamiltonian of the composed system is $H_T = H_A + H_R + g(T_A) H_R P_E$. Then, in a similar manner Eqs. \eqref{eq:schrod-general} and \eqref{eq:general-2clocks} were derived in the previous section, we obtain the Schr\"odinger equation
\begin{equation}
    i\hbar \frac{\partial}{\partial t_A} |\psi(t_A)\rangle = \left[H_R + g(t_A) H_R P_E\right] |\psi(t_A)\rangle,
\end{equation}
where $|\psi(t_A)\rangle \equiv \langle t_A|\Psi\rangle\rangle$. This means that, from the perspective of clock $A$, the evolution is generated by the effective Hamiltonian
\begin{equation}
    H_1 \equiv H_R + g(t_A) H_R P_E,
\end{equation}
which is the usual Hamiltonian of a time-independent system during a measurement. This is expected since, in standard quantum mechanics, time is an external parameter, as it is in this case.

Using the Heisenberg equation of motion to study the evolution of $T_B$ with respect to clock $A$, we conclude that
\begin{equation}
    \frac{d}{dt_A}T_B = -\frac{i}{\hbar} [T_B,H_1] = I+g(t_A)P_E.
    \label{eq:ext-perspective-a}
\end{equation}

This shows that, when $g$ vanishes, the flow of time in both clocks is the same. In fact, in this case, from the perspective of clock $A$, the variance of $T_B$ at any instant originates from its initial variance. In particular, if both clocks start localized and synchronized, they remain localized and synchronized.

Yet, clock $B$ ticks faster than clock $A$ whenever $g$ is non-null, i.e., during the measurement of energy. If $g$ is a function whose integral over time grows smoothly, then the transition to a faster ticking also happens smoothly. However, this is not always the case. A dramatic example can be observed by assuming a highly idealized case where $g$ is the delta function. In this case, from the perspective of clock $A$, there is a sudden ``jump'' of the pointer of clock $B$.

While the dependence of $T_B$ on $P_E$ might seem surprising, it reflects the fact that clock $B$ interacts with the pointer. However, not every interaction with a clock leads to a change in its flow of time. To understand the change observed in the case of interest here, it suffices to analyze the term $g(t_A) H_B P_E$ from the von Neumann interaction. Although it is usually thought of as a term that causes a shift (generated by $P_E$) of a quantity associated with $H_B$ in the measurement device's pointer, it can be also understood as a term causing a shift (generated by $H_B$) of a quantity associated with $P_E$ in the pointer (i.e., time) of clock $B$. Generally, by this reasoning, interactions with clocks intermediated by their Hamiltonian should affect their flow of time by quantities associated with the interacting system.

\begin{figure}
    \centering
    \includegraphics[width=\columnwidth]{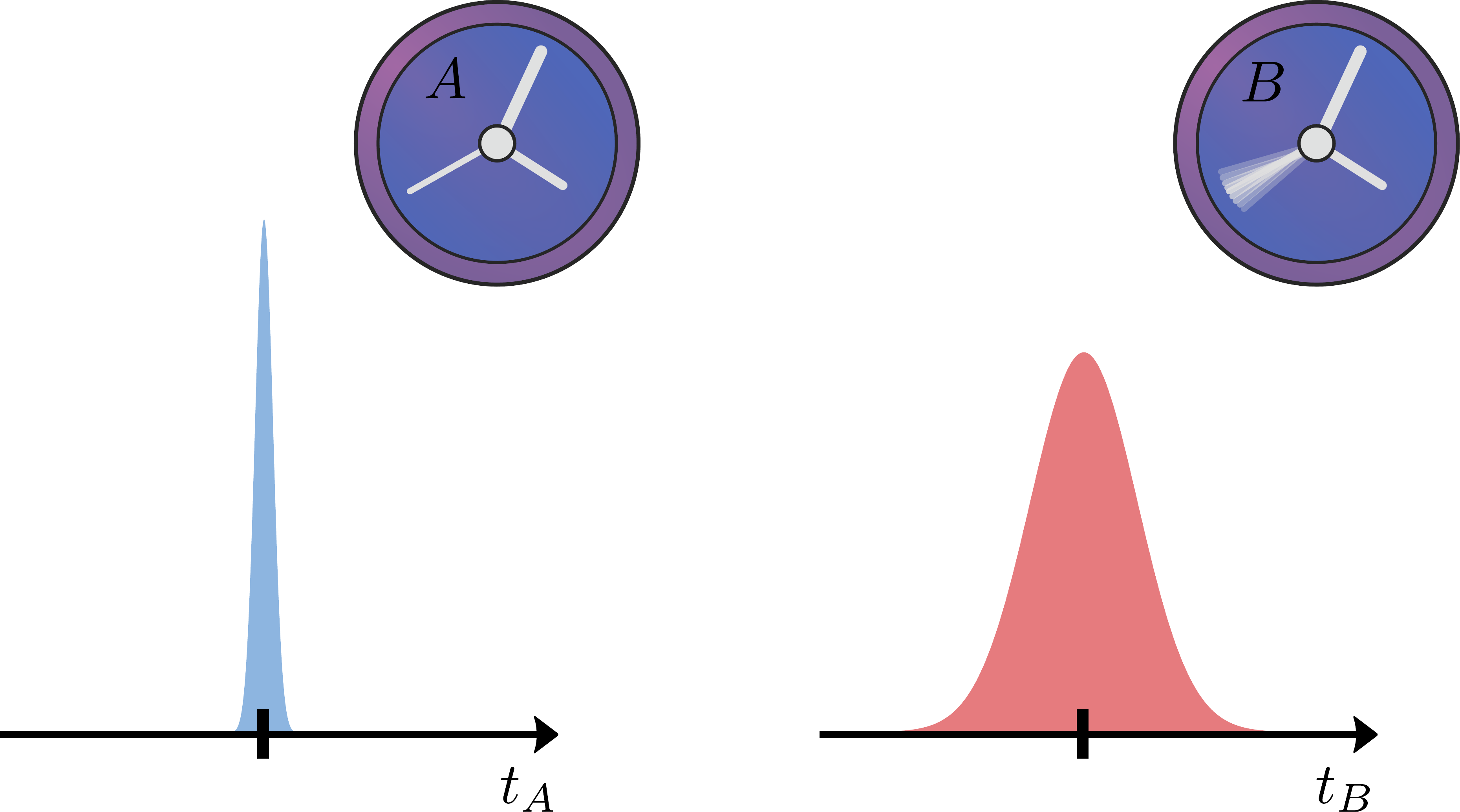}
    \caption{Localizability of events with multiple clocks. A well-localized event in clock $A$ has a variance in clock $B$, and vice-versa.}
    \label{fig:clocks-rel}
\end{figure}

Now, using Eq. \eqref{eq:constraint} and defining $|\phi(t_B)\rangle \equiv \langle t_B|\Psi\rangle\rangle$, we obtain the following Schr\"odinger equation from the perspective of clock $B$:
\begin{equation}
    \begin{aligned}
        i\hbar [I+g&(T_A)P_E] \frac{\partial}{\partial t_B} |\phi(t_B)\rangle = H_A |\phi(t_B)\rangle \\
             &+\left[I+g(T_A)P_E\right] [H_S + H_{int}(t_B)] |\phi(t_B)\rangle.
    \end{aligned}
    \label{eq:redshift-ext}
\end{equation}
Moreover, recalling that $P_E$ is assumed to only take non-negative values, which implies that $I+g(T_A)P_E$ is invertible, the effective Hamiltonian is
\begin{equation}
    H_2 \equiv \left[I+g(T_A)P_E\right]^{-1} H_A + H_S + H_{int}(t_B).
\end{equation}
Observe that the function $g$ that controls the measurement continues to be a function of the operator $T_A$, while it was a function of the parameter $t_A$ when considering $A$'s perspective. In a scenario like this, a well-localized event in clock $A$ has a variance from clock $B$'s perspective, a result illustrated in Fig. \ref{fig:clocks-rel} and previously discussed, for instance, in Ref. \cite{castro2020quantum}. This means, in particular, that the start and end of the measurement, which are well-localized in $A$, have variance in $B$.

Moreover, in a sense, Hamiltonian $H_2$ corresponds to a measurement of $H_A$. However, the measurement function $g$ is controlled by the time in clock $A$, which is an observable from the perspective of $B$. Because of it, the measurement Hamiltonian would have been symmetrized in more standard treatments of the process, which is not the case here. In fact, $H_2$ is non-Hermitian and, as a consequence, the evolution of clock $A$ from the perspective of clock $B$ is non-unitary, a characteristic that will be further discussed later in this work. For now, observe that the Heisenberg equation of motion for $T_A$ with respect to $t_B$ gives
\begin{equation}
    \frac{d}{dt_B}T_A = -\frac{i}{\hbar} [T_A,H_2] = [I+g(T_A)P_E]^{-1}.
    \label{eq:ext-perspective-b}
\end{equation}
This result is expected in view of Eq. \eqref{eq:ext-perspective-a} since it gives the inverse of the passage of time on clock $B$ with respect to clock $A$. However, once again, it should be noticed that here the argument of $g$ is an operator, and no longer a parameter. Then, with respect to clock $B$, if $T_A$ starts with a variance, the duration of the measurement will have a variance as well. Nevertheless, on average, the ``flow of time'' during the measurement of energy is expected to be smaller in clock $A$, from the perspective of both clock $A$ and clock $B$.

Even though Eq. \eqref{eq:ext-perspective-b} shows consistency across different clock perspectives, it raises questions about the use of the Heisenberg equation here. While Schr\"odinger's and Heisenberg's representations are unitarily equivalent, since the evolution from clock $B$'s perspective is, in general, non-unitary, it is not trivial to obtain the Heisenberg equation from the effective Hamiltonian in the Schr\"odinger equation. Nevertheless, this can be done with the introduction of an ``indefinite metric'' in the Hilbert space, a method introduced by Dirac \cite{dirac1942bakerian} and briefly explained in Appendix \ref{app:heisenberg}.

\subsection{Measurement carried out by an internal system}

This subsection considers, once more, the total energy measurement of system $R$, although, this time, the measurement in question is carried out by an internal system. A concrete example of this type of measurement of total energy of an isolated system was given by Aharonov and Reznik \cite{aharonov2000weighing}. They considered an isolated planet with a large radius $R$ ejecting a tiny portion $m$ of its total mass radially outwards with a known speed. Then, using the relativistic correspondence of mass and energy, they showed that it is possible to infer the total energy of the planet by measuring how long it takes for $m$ to fall back.

\begin{figure}
    \centering
    \includegraphics[width=\columnwidth]{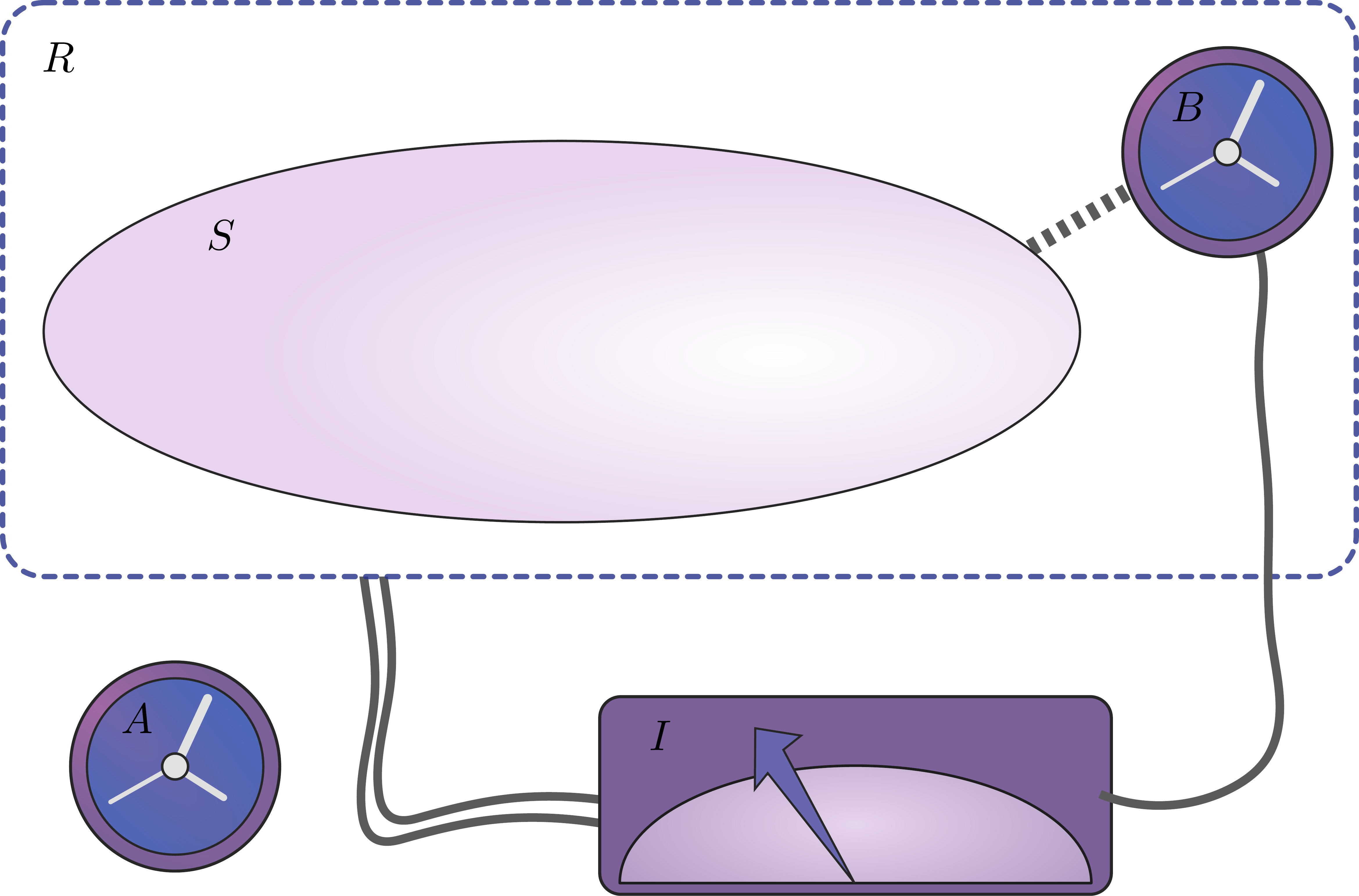}
    \caption{Representation of the relevant parts in a measurement of energy carried out by an internal part. In this case, the internal apparatus $I$ measures the energy of the remaining parts, i.e., system $R$, composed of clock $B$ and system $S$. Clock $A$ is an external clock, which does not interact with any other system during this process. For the measurement, the internal clock $B$ is used as a reference for time.}
    \label{fig:int-measurement}
\end{figure}

Here, we use an abstract model for these measurements, similarly to what was done in the previous subsection. We assume that system $R$ has access to an apparatus $I$, and the history state $|\Psi\rangle\rangle$ is an element of $\mathcal{H}_A \otimes \mathcal{H}_R \otimes \mathcal{H}_I$, as represented in Fig. \ref{fig:int-measurement}. Note that the apparatus may be assumed to be an internal degree of freedom of the system whose energy is left out of the measurement.

Similarly to what was done in the previous section, the free evolution of the pointer $I$ will be neglected. As a result, the measurement can be represented by the following von Neumann interaction
\begin{equation}
    H_{VN} \equiv \frac{1}{2}[g(T_B) H_R + H_R g(T_B)] P_I,
\end{equation}
where $H_R=H_B+H_S+H_{int}(T_B)$, $P_I$ is the conjugate momentum of apparatus $I$, presupposed to only take non-negative values, and $g$ is the same function introduced in the previous section. Observe the necessary symmetrization of the product $g(T_B)H_R$ due to the lack of commutativity between $T_B$ and $H_B$, as previously considered, for instance, in Ref. \cite{massar2005measurement}. Also, we assume that, in clock $B$, the measurement starts at $t_B=0$ and ends at $t_B=\tau$.

Since the total Hamiltonian of the composed system is $H_T = H_A + H_R + H_{VN}$, in a similar manner Eqs. \eqref{eq:schrod-general} and \eqref{eq:general-2clocks} were derived in the previous section, we obtain the Schr\"odinger equation
\begin{equation}
    i\hbar \frac{\partial}{\partial t_A} |\psi(t_A)\rangle = H_3 |\psi(t_A)\rangle,
\end{equation}
where $|\psi(t_A)\rangle \equiv \langle t_A|\Psi\rangle\rangle$ and
\begin{equation}
    H_3 \equiv H_R + \frac{1}{2}[g(T_B) H_R + H_R g(T_B)] P_I
\end{equation}
is the effective Hamiltonian from the perspective of clock $A$. With that, the Heisenberg equation of motion for $T_B$ with respect to $t_A$ is
\begin{equation}
    \frac{d}{dt_A}T_B = -\frac{i}{\hbar} [T_B,H_3] = I+g(T_B)P_I.
    \label{eq:int-perspective-a}
\end{equation}

On the other hand, using Eq. \eqref{eq:constraint} and defining $|\phi(t_B)\rangle \equiv \langle t_B|\Psi\rangle\rangle$, we obtain
\begin{equation}
    \begin{aligned}
        i\hbar &[I+g(t_B)P_I] \frac{\partial}{\partial t_B} |\phi(t_B)\rangle = \\
        &=\left[H_A -\frac{i\hbar}{2} g'(t_B) P_I\right] |\phi(t_B)\rangle \\
            & \ \ \ +[I+g(t_B)P_I] [H_S + H_{int}(t_B)] |\phi(t_B)\rangle,
    \end{aligned}
    \label{redshift-int}
\end{equation}
where it was used the fact that $I+g(t_B)P_I$ can be inverted. This means that, from the perspective of clock $B$, the evolution is generated by the effective Hamiltonian
\begin{equation}
    \begin{aligned}
        H_4 \equiv &[I+g(t_B)P_I]^{-1} \left[ H_A -\frac{i\hbar}{2} g'(t_B) P_I\right] \\
        &+ H_S + H_{int}(t_B),
    \end{aligned}
\end{equation}
which, similarly to $H_2$, is a non-Hermitian operator.

Then, using the Heisenberg equation of motion to study the evolution of $T_A$ with respect to $t_B$, we conclude that
\begin{equation}
    \frac{d}{dt_B}T_A = -\frac{i}{\hbar} [T_A,H_4] = [I+g(t_B)P_I]^{-1},
    \label{eq:int-perspective-b}
\end{equation}
which, as expected, up to the difference that the argument of $g$ is, now, a parameter, the inverse of the relation between the passage of time in the two clocks given by Eq. \eqref{eq:int-perspective-a}. Here, again, the use of the Heisenberg equation is justified with the introduction of an indefinite metric in the Hilbert space, as discussed in Appendix \ref{app:heisenberg}.

\section{Time-energy uncertainty relations}
\label{sec:uncertainty}

Now, we shall discuss how the comparison between the flow of interior and the exterior time during an energy measurement provides insights into time-energy uncertainty relations associated with the von Neumann measurement of energy. As already mentioned, knowing the Hamiltonian beforehand makes a difference when studying these relations since, otherwise, the process of effectively measuring the Hamiltonian includes a preliminary step of estimating it \cite{aharonov2002measuring}. Here, we focus on the cases the Hamiltonian of system $R$ is known and leave the study of estimation of Hamiltonians to future work.

It should be noticed that, because of the explicit inclusion of the frames associated with the internal and external clocks, our analysis becomes slightly more subtle than others presented up until now. While, like the previous works, we can consider which time (internal or external) one wants to optimize when an internal or external system carries out the measurement, we have to also specify with respect to which clock frame (internal or external) the question is being asked. However, our approach has the advantage of introducing a unified framework to analyzing time-energy uncertainty relations, which qualifies us to recover previous results and determine the existence or absence of such relations in scenarios where it is still unknown.

\subsection{Measurement carried out by an external system}

First, we consider measurements carried out by an external system from the perspective of $A$. Observe that the measurement function $g$ is controlled by clock $A$ and, then, nothing in the von Neumann measurement procedure we follow imposes a limit on how small the interval where $g$ does not vanishes is, i.e., how fast the measurement can be performed. As a consequence, the external time can be arbitrarily minimized without requiring any compromising of the measurement accuracy, i.e., without imposing any restriction on the constant $K$. This corresponds to the result obtained by Aharonov and Bohm in Ref. \cite{aharonov1961time}, which shows the nonexistence of an uncertainty relation in this scenario.

\begin{figure}
    \centering
    \includegraphics[width=\columnwidth]{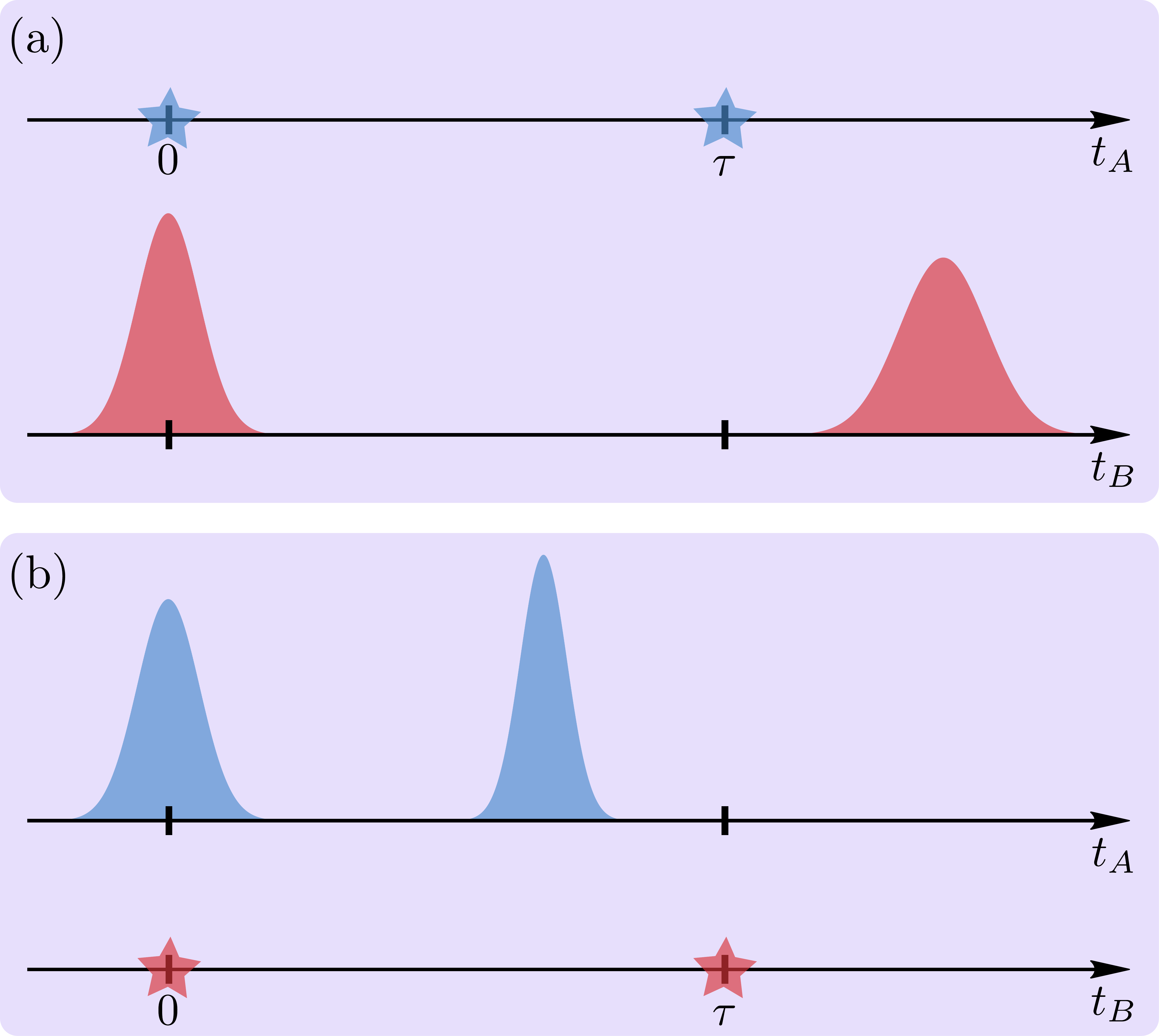}
    \caption{Representation of the duration of energy measurements in internal and external clocks. Clock $A$ is external and clock $B$ is internal to the measured system. (a) When an external system carries out the measurement, its duration in $B$ is longer on average. (b) However, when an internal part of the system is responsible for the measurement, its duration in the external clock is shorter on average.}
    \label{fig:measurement}
\end{figure}

Now, a question that has not been dealt with in the literature thus far concerns the possibility of optimizing the internal time in this scenario. For that, observe that Eq. \eqref{eq:ext-perspective-a} lead to
\begin{equation}
    T_B(\tau) - T_B(0) = \tau I + K P_E.
\end{equation}
This result is represented in Fig. \ref{fig:measurement}a. It shows that the internal duration of the measurement decreases with $\tau$, i.e., with the external duration, which can, in principle, be as small as desirable. However, this can only reduce the internal duration up to $KP_E$. Thus, in order to obtain an arbitrarily small internal duration, it is necessary to decrease the value of the constant $K$. More precisely,
\begin{equation}
    \langle T_B(\tau) - T_B(0) \rangle \geq K \langle P_E\rangle>0.
    \label{eq:ineq}
\end{equation}
Here, the inequality refers to statistics associated with repeated implementations of the measurement protocol. The average on the left-hand side refers to the average duration of the experiment in clock $B$. On the other part of the inequality, $\langle P_E\rangle$ is the average value of the conjugate momentum of the pointer $E$.

Inequality \eqref{eq:ineq} means that there exists a trade-off between the precision of the measurement (which is associated with $K$) and how long the measurement lasts in the internal clock, i.e., there exists a time-energy uncertainty relation in the sense of Landau and Peierls \cite{landau1931erweiterung} discussed in the Introduction, i.e., quantum mechanics imposes a minimum duration for measurements of energy with a desired precision. Note that such a relation is independent of the variance of the operator $T_B$. Also, the existence of such an inequality cannot be attributed to disturbances on clock $B$ caused by the measurement since they vanish, as explained in Appendix \ref{app:disturbance}.

Moreover, taking variances into consideration, we obtain a different type of uncertainty relation. In fact, observe that the variance $\Delta T_B$ in the duration of the experiment in clock $B$ from clock $A$'s perspective is associated with the variance $\Delta P_E$ of $P_E$. More precisely, $\Delta T_B$ grows with $\Delta P_E$. Then, in order to obtain a smaller $\Delta T_B$, it is necessary to decrease $\Delta P_E$. However, $\Delta P_E$ is also associated with the precision of the measurement since a smaller $\Delta P_E$ implies that pointer of the measurement device is more delocalized. As a result, a smaller $\Delta T_B$ leads to an increase in the precision of the measurement, implying an uncertainty relation of the type $\Delta E \Delta T_B \geq 1$, where $\Delta E$ is the uncertainty of the energy measurement.

Now, from the perspective of clock $B$, the analysis of the measurement carried out by an external system leads mostly to the same conclusions. In fact, with some probability, we can say that the measurement starts in a certain instant $t^B_i$ and ends at $t^B_f$ in clock $B$. Then, Eq. \eqref{eq:ext-perspective-b} implies that
\begin{equation}
    t^B_f - t^B_i = \int_{t^B_i}^{t^B_f} [I+g(T_A(t_B))P_E] \, \frac{dT_A}{dt_B}(t_B) \, dt_B
\end{equation}
or, equivalently,
\begin{equation}
    t^B_f - t^B_i = T_A(t_f^B) - T_A(t_i^B) + K P_E.
\end{equation}
Note that, on average, $T_A(t_f^B) - T_A(t_i^B)$ gives $\tau$. Then, the only difference in the analysis from $B$'s perspective is that the time in clock $A$ is given by an operator and, hence, has a variance associated with it, which, in turn, leads to a variance in the duration of the measurement. This leads to an uncertainty relation for the duration of the measurement and its precision associated with the variances of both clocks $A$ and $B$. However, in terms of the uncertainty relations of main interest here, conclusions about their existence or not are independent of perspective.

\subsection{Measurement carried out by an internal system}

We consider now the measurement carried out by an internal system from the perspective of clock $B$. Again, we start by pointing out that nothing in our description disallows the measurement to be conducted arbitrarily fast. Then, there exists no uncertainty relation in this case, a conclusion that corroborates the result obtained by Massar and Popescu in Ref. \cite{massar2005measurement}.

Furthermore, still from the perspective of clock $B$, we can address the question about the existence of such a relation for the external time, which, to the best of our knowledge, has not been answered in the literature prior to this work. Note that Eq. \eqref{eq:int-perspective-b} implies that
\begin{equation}
    T_A(\tau) - T_A(0) = \int_0^\tau [I+g(t_B)P_I]^{-1} dt_B,
\end{equation}
as represented in Fig. \ref{fig:measurement}b. With the hypothesis that the evaluation of $P_I$ is non-negative, it holds that the norm of the last integral is bounded by $\tau$ (independently of $K$), i.e., the the duration of the measurement in clock $A$ can be as small as its duration in clock $B$ --- in fact, it is, in
general, smaller since the evaluation of $I+g(t_B)P_I$ is always greater or equal one. Considering that the duration can be arbitrarily small in clock $B$, there exists no time-energy uncertainty relation for clock $A$ from the perspective of clock $B$.

This is a rather surprising result. In fact, Massar and Popescu briefly consider this scenario in Ref. \cite{massar2005measurement}, where they presented a conjecture stating that an uncertainty relation should hold.

Before jumping into $A$'s perspective, it should be noted once again the existence of an uncertainty relation for the duration of the measurement in clock $A$ associated with the variance of $T_A$.

Now, from the perspective of clock $A$, the conclusions are, once again, similar. Suppose the measurement starts and ends at $t_i^A$ and $t_f^A$, respectively. Then, Eq. \eqref{eq:int-perspective-a} implies that
\begin{equation}
    t_f^A - t_i^A = \int_{t_i^A}^{t_f^A} [I+g(T_B(t_A))P_I]^{-1} \, \frac{dT_B}{dt_A}(t_A) \, dt_A.
\end{equation}
Since the evaluation of $g(T_B)P_I$ is assumed to be non-negative, the average of the right-hand is not greater than the average of the integral of $dT_B/dt_A$, which is $\tau$. Hence, the only difference in the analysis from $A$'s frame is the variance associated with $T_B$, which leads to an uncertainty relation for the duration of the measurement in both clocks $A$ and $B$. Here, again, it should be noted that disturbances on clock $B$ do not affect this analysis since they can be taken to be null, as discussed in Appendix \ref{app:disturbance}.

\section{Emergence of non-unitarity}
\label{sec:nonunitarity}

As noticed earlier in the context of energy measurements, an interesting feature of the dynamics generated by the measurement of energy is that, from the perspective of the internal clock, it is non-unitary regardless of whether an external or an internal system carries out the measurement.

In relativistic theories, energy and mass are tightly related. Then, while measuring the energy of a system that includes an internal clock, there is a sense in which the mass of the system and, in particular, the mass of the clock is being weighed. Could, then, the non-unitarity observed in this work be related to gravitational effects? After all, non-unitarity is expected to play a role in quantum gravity, see e.g. \cite{hawking1982unpredictability, unruh1995evolution, penrose1996gravity, gambini2007fundamental}. However, gravitationally interacting clocks have been considered in the literature \cite{ruiz2017entanglement, smith2019quantizing, castro2020quantum}, and it was shown that they lead to unitary dynamics. In fact, in these studies, the interaction term between two clocks $A$ and $B$ was of the type $\lambda H_A H_B$. Then, it can be checked that the effective Hamiltonian of $B$ with respect to $A$ is $(I+\lambda H_B)^{-1} H_B$. Similarly, the effective Hamiltonian of $A$ with respect to $B$ is $(I+\lambda H_A)^{-1} H_A$. Since these two effective Hamiltonians are Hermitian, the evolution generated by them is unitary.

Then, why is the result different here? The answer to this question lies in the fact that the von Neumann interaction includes products of the time operator of a clock by the individual Hamiltonian of the same or other clocks. In fact, if such a condition is matched, the dynamics of the clocks associated with the time operator in the interaction is non-unitary with respect to the clock whose individual Hamiltonian appears multiplying it. To see that, consider a system composed of $n$ clocks $C_1, C_2, \hdots, C_n$ and let the total Hamiltonian be
\begin{equation}
    H_T = \sum_k H_{C_k} + f(T_{C_{r_1}}, \hdots, T_{C_{r_m}}) H_{C_s} + H_{int},
\end{equation}
where $m<n$, the indices $r_1, r_2, \hdots, r_m$,  and $s$ are different elements of $\{1, 2, \hdots, n\}$, and $H_{int}$ is not a function of $H_s$. Then, from the perspective of $C_s$ we can write
\begin{equation}
    \begin{aligned}
        i\hbar [I + f(&T_{C_{r_1}}, \hdots, T_{C_{r_m}})] \frac{\partial}{\partial t_{C_s}} |\psi(t_{C_s})\rangle = \\
             &=\left[\sum_{k\neq s} H_{C_k} + H_{int}\right] |\psi(t_{C_s})\rangle,
    \end{aligned}
    \label{eq:persp-cs}
\end{equation}
i.e., the effective Hamiltonian, which is to be compared with $H_2$, is
\begin{equation}
    H_{eff}^{C_s} \equiv [I + f(T_{C_{r_1}}, \hdots, T_{C_{r_m}})]^{-1} \left[\sum_{k\neq s} H_{C_k} + H_{int}\right],
\end{equation}
assuming that $I + f(T_{C_{r_1}}, T_{C_{r_2}}, \hdots, T_{C_{r_m}})$ is invertible. This Hamiltonian is manifestly non-Hermitian. More specifically, the parts associated with clocks $C_{r_1}, C_{r_2}, \hdots, C_{r_m}$ are non-Hermitian and, hence, have non-unitary evolution because of the lack of commutativity between their individual time and Hamiltonian operators.

The only scenario that remains to be analyzed now corresponds to the case where an interaction between the time operator and the individual Hamiltonian of the same clock exists. For that, suppose the total Hamiltonian is given by
\begin{equation}
    \begin{aligned}
        H_T &= \sum_k H_{C_k} + \frac{1}{2} [f(T_{C_s}) H_{C_s} + H_{C_s} f(T_{C_s})] + H_{int} \\
        &= \sum_k H_{C_k} + f(T_{C_s}) H_{C_s} - \frac{i\hbar}{2} f'(T_{C_s}) +  H_{int},
    \end{aligned}
\end{equation}
where, once again, $H_{int}$ is assumed to not depend on $H_{C_s}$. From clock $C_s$'s perspective, this leads to
\begin{equation}
    \begin{aligned}
        i\hbar [I + &f(t_{C_s})] \frac{\partial}{\partial t_{C_s}} |\psi(t_{C_s})\rangle = \\
        &=\left[\sum_{k\neq s} H_{C_k} - \frac{i\hbar}{2} f'(t_{C_s}) +  H_{int}\right] |\psi(t_{C_s})\rangle,
    \end{aligned}
    \label{eq:time-hamiltonian}
\end{equation}
which shows that the effective Hamiltonian is
\begin{equation}
    H_{eff}^{C_s} = [I + f(t_{C_s})]^{-1} \left[\sum_{k\neq s} H_{C_k} - \frac{i\hbar}{2} f'(t_{C_s}) +  H_{int}\right],
\end{equation}
assuming that $I + f(t_{C_s})$ is invertible. This Hamiltonian is be compared with $H_4$. Although there is no commutation problem, the Hamiltonian in this case is also non-Hermitian because of the imaginary unit multiplying $f'(t_{C_s})$.

Despite the differences between the discussion in the present work and the gravitationally interacting clocks studied in Ref. \cite{castro2020quantum}, there exists an aspect that leads to a similar analysis. In Ref. \cite{castro2020quantum}, factors like $(I+\lambda H_A)^{-1}$ and $(I+\lambda H_B)^{-1}$ were interpreted as ``redshift'' operators. Also, these operators were combined with the time derivative present in their respective Schr\"odinger equation, resulting in derivatives with respect to new time operators: $\partial/\partial \hat{t}_A \equiv (I+\lambda H_B) \partial/\partial t_A$ and $\partial/\partial \hat{t}_B \equiv (I+\lambda H_A) \partial/\partial t_B$. In a similar manner, the term between brackets on the left-hand side of Eq. \eqref{eq:persp-cs} can be interpreted as a sort of ``redshift'' operator, which, combined with the $t_{C_s}$-derivative operator, results in a derivative with respect to a time operator $\hat{t}_s$. This allows us to write
\begin{equation}
    i\hbar \frac{\partial}{\partial \hat{t}_s} |\psi\rangle = \left[\sum_{k\neq s} H_{C_k} + H_{int}\right] |\psi\rangle.
\end{equation}
Observe that, with these coordinates, the unitarity of the dynamics is restored. However, a complete change to clock $C_s$'s referential leads to the observation of the ``redshift,'' an analogy that leads to the conclusion that clock $C_s$ is not an inertial frame of reference.

A similar analysis can be made with Eq. \eqref{eq:time-hamiltonian}. However, in this case, the non-unitarity of the dynamics persists even when the derivative with respect to an operator $\hat{t}_s$, which consists of the $t_{C_s}$-derivative operator combined with the ``redshift'' factor, is considered.

\section{Discussion}
\label{sec:conclusion}

Using the Page and Wootters framework, we have studied how the relation between internal and external ``flow of time'' in quantum clocks changes during energy measurements performed either by an external or internal system. Some of these measurements, as was discussed, cause disturbances in the energy of the systems, which are independent of the duration of the measurements. While there is no penalty for performing a fast energy measurement (in terms of disturbances added to it), the change in the flow of time between an external and an internal clock reveals a minimum duration in the internal clock of a measurement with a given precision carried out by an external system. Any other scenario considered here led to the conclusion of an absence of a time-energy uncertainty relation in the sense of Landau and Peierls \cite{landau1931erweiterung}.

Importantly, the time-energy uncertainty relations considered here assume the Hamiltonian being measured to be known beforehand. However, as mentioned in the Introduction, if this is not the case, the Hamiltonian has to be estimated. While Ref. \cite{aharonov2002measuring} shows the existence of a minimum duration in an external clock for an external system to estimate the Hamiltonian of the system of interest, a discussion about the relation of this time interval and the interval observed in an internal clock remains to be conducted. Moreover, the scenario where an internal system attempts to estimate its total Hamiltonian has yet to be investigated. An interesting aspect of the proof provided by Aharonov, Massar, and Popescu in Ref. \cite{aharonov2002measuring} is that it involves the time necessary for the state of a system to become orthogonal to its initial state under a designed evolution. Then, the type of uncertainty relation that arrives from it is of the geometric type, briefly mentioned in the Introduction and originally discussed by Anandan and Aharonov in Ref. \cite{anandan1990geometry} --- followed by Margolus and Levitin in Ref. \cite{margolus1998maximum}.

While we concentrated on the measurement's unitary interaction in this work, it is well-known that von Neumann's description of a measurement includes a ``collapse'' (or post-selection) in case the pointer is found in a superposition at the end of the process. The inclusion of this final step in our analysis adds various new subtleties. On the one hand, since time and energy of a clock are canonically conjugated, there exists an uncertainty relation for a measurement carried out by an internal system, meaning that time in the internal clock can only be known up to a certain precision given a desired precision for the measurement of energy. This is a Heisenberg-like uncertainty relation and is captured by entropic time-energy uncertainty relations recently introduced by Boette \textit{et al.} \cite{boette2016system} and Coles \textit{et al.} \cite{coles2019entropic}. On the other hand, the inclusion of a ``collapse'' in our description would necessarily violate Eq. \eqref{eq:constraint}. This is a known problem in the measurement of quantum clocks and, more generally, of any quantum system satisfying the Wheeler-DeWitt equation. A typical solution to this problem consists of the inclusion of ancilla systems that act as the measurement device and register the measurement outcome \cite{hellmann2007multiple, giovannetti2015quantum, castro2020quantum, trassinelli2021conditional}. This perspective solves at least the operational problem of computing the probability of the outcome of (possibly many) measurements. Since our approach includes measurement devices, this problem has been, then, already dealt with (at least from this perspective). A different approach consists of updating the state with a ``collapse'' and, then, applying a map to transform the resulting state into a state that satisfies the Wheeler-DeWitt equation. This procedure, however, cannot be formulated in a perspective-neutral manner \cite{yang2020switching}. We have not addressed this approach here.

Observe that the system $S$ introduced in Section \ref{sec:measurement} served exclusively to enlarge the generality of our measurement model, not affecting the time-energy uncertainty relations nor the non-unitarity of the evolution. This is the case because the interaction between system $S$ and clock $B$ was assumed to be mediated by $T_B$. If interactions mediated by $H_B$ were allowed, they would be given by a term that contains the product $O_S H_B$, where $O_S$ is an observable of $S$. This, in turn, causes a shift (generated by $H_B$) associated with $O_S$ in the pointer (i.e., time) of clock $B$, as discussed in Section \ref{sec:measurement}. Although this effect alone would not necessarily lead to non-unitary dynamics, it could in some cases. Further analysis of generic interactions that lead to non-unitarity is left for a future work. Here, for simplicity, as already stated, we have restricted our study to interactions between $S$ and $B$ modeled by a function of $T_B$ on the clock space.

Finally, there still exist many remaining questions concerning the emergence of non-unitarity observed in our study. An immediate analysis can attribute it to the asymmetry imposed by the von Neumann measurement, as it was already discussed throughout the text. However, since this type of evolution always concerns the evolution of the external clock from the perspective of the internal one, it may be connected with the decoherence of the internal clock during the measurement. This is a perspective that deserves further examination. More precisely, one may investigate the connection between non-unitarity emergent in this work with decoherence and, in particular, with Refs. \cite{gambini2007relational} and \cite{pikovski2015universal}, where time dilation was associated with decoherence. Moreover, following the approach of Refs. \cite{martinelli2019quantifying} and \cite{carmo2021quantifying}, the concepts of \textit{shared} and \textit{mutual asymmetry} may shed some light on the non-unitary dynamics manifested in this work. Also, it could be that the clock is not an inertial frame of reference. In this case, deviations of the metric $\eta$ discussed in Appendix \ref{app:heisenberg} from the identity could be used as an indicator of non-inertiality.

\acknowledgements{We thank the anonymous referees for the thorough reading of this work and the constructive comments. We are also grateful to Yakir Aharonov, Renato Moreira Angelo, Pedro Ruas Dieguez, Cristhiano Duarte, Luis Pedro Garc\'ia-Pintos, and Marcin Nowakowski for helpful discussions and constructive remarks regarding this text. This research was supported by grant number FQXi-RFP-CPW-2006 from the Foundational Questions Institute and Fetzer Franklin Fund, a donor-advised fund of Silicon Valley Community Foundation, by the Israeli Innovation authority (grants 70002 and 73795), by the Pazy foundation, and by the Quantum Science and Technology Program of the Israeli Council of Higher Education.}

\bibliographystyle{plain}

\onecolumn\newpage

\appendix

\section{Non-unitarity and the Heisenberg picture}
\label{app:heisenberg}

Although most of our analysis is centered around the unitary dynamics given by the external clock $A$, the dynamics --- and, in particular, the Heisenberg equation --- from the perspective of the internal clock $B$ was used to show some level of consistence across the different perspectives. However, the Heisenberg and the Schr\"odinger pictures are unitarily equivalent and one may question whether the use of the Heisenberg equation in the non-Hermitian scenario is valid. The main goal of this appendix is to justify this use.

When analyzing non-Hermitian Hamiltonians, like the effective ones from clock $B$'s perspective found in Section \ref{sec:measurement}, a method introduced by Dirac \cite{dirac1942bakerian} and further studied, for instance, in Refs. \cite{pauli1943dirac, lee1969negative, scholtz1992quasi} comes in handy. It entails the replacement of the usual inner product in the Hilbert space by a product associated with an indefinite Hermitian metric $\eta$, i.e.,
\begin{equation}
    \langle\psi|\phi\rangle_\eta \equiv \langle\psi|\eta|\phi\rangle.
\end{equation}
The spectrum of $\eta$ is simply composed of $-1$ and $1$.

Then, in addition to the Hermitian conjugate $O^\dag$ of an operator $O$, it is also possible to define its conjugate $O^*$ as
\begin{equation}
    O^* \equiv \eta^{-1} O^\dag \eta.
\end{equation}
Observables that are self-adjoint in this sense are called \textit{$\eta$-pseudo-adjoint}. Physical observables and, in particular, the non-Hermitian Hamiltonian of interest $H$, must be $\eta$-pseudo-adjoint. With that, it can be observed that the $\eta$-norm of a state vector is kept constant throughout the evolution generated by $H$. As a result, the expected value
\begin{equation}
    \langle O\rangle = \langle\psi|O|\psi\rangle_\eta = \langle\psi|\eta O|\psi\rangle
\end{equation}
of a non-explicitly time-dependent operator $O$ is such that
\begin{equation}
    \frac{d}{dt}\langle O\rangle = -\frac{i}{\hbar} \langle [O,H]\rangle,
\end{equation}
which allows us to recover the Heisenberg picture.

It should be emphasized that the metric $\eta$ diverges from the identity only if the Hamiltonian is non-Hermitian, like Hamiltonians $H_2$ and $H_4$. As a result, Hamiltonians $H_1$ and $H_3$ lead to the familiar structure of quantum mechanics.

The main concern about the introduction of indefinite metrics that takes into consideration some symmetries of the Hamiltonian is the fact that it typically introduces ghost states, i.e., states with negative norm, which lacks a clear physical interpretation \cite{pauli1943dirac}. However, it could be the case that the system has some extra symmetry and, when this symmetry is considered, the metric becomes positive-definite, which transforms the ghost states into standard states, i.e., states with positive norm.

A particular example of the above is given by $\mathcal{PT}$-symmetric Hamiltonians, which are invariant under the composition of parity inversion time reversal \cite{bender1998real} --- see also \cite{lee1954some, wu1959ground, brower1978critical, fisher1978yang}. The former is represented by the parity operator $\mathcal{P}$ and the latter by the time reversal operator $\mathcal{T}$. A remarkable characteristic of $\mathcal{PT}$-symmetric Hamiltonians is that their eigenvalues are real. Now, if we consider the conjugation with respect to the indefinite metric $\mathcal{PT}$, the resultant theory contains ghost states. This is a problem that puzzled many in the community for decades. However, it was later observed that systems with unbroken $\mathcal{PT}$-symmetry have a third symmetry $\mathcal{C}$ such that the resulting $\mathcal{CPT}$ metric is positive-definite \cite{bender2002complex}.

While Hamiltonians $H_2$ and $H_4$ are, in general, not $\mathcal{PT}$-symmetric and their induced metric may lead to ghost particles \textit{a priori}, one may wonder if an extended analysis could lead to the introduction of a positive-definite metric given a measurement function $g$. If this turns out to indeed be the case, there will be a sense in which the resultant evolution is still unitary.

\section{Measurement and disturbance of the system's dynamics}
\label{app:disturbance}

In this appendix, we study disturbances on system $R$, and in particular on clock $B$, caused by measurements of energy. The goal is to show that they do not influence the time-energy uncertainty relations analyzed in Section \ref{sec:uncertainty}.

To start, we observe that the standard von Neumann measurement of energy does not disturb the evolution of the observable of interest of the measurement. This can be seen with the measurement of $H_R$ given by $H_1$, which corresponds to the standard treatment of the measurement in quantum mechanics. In fact, it can be checked that the $t_A$-derivative of $H_R$ vanishes and, as a consequence, $[dH_R/dt_A,H_R]=0$, which means that the evolution of $H_R$ is not disturbed by the measurement interaction.

Moving on, we study whether the measurement of energy carried out by the internal system disturbs system $R$'s dynamics from the external clock's perspective. In this regard, it can be obtained that
\begin{equation}
    \frac{d}{dt_A} H_R = -\frac{i}{\hbar} [H_R, H_3] = -\frac{1}{2} \left\{g'(T_B),H_R\right\} P_I.
    \label{eq:hr-change}
\end{equation}
and
\begin{equation}
    \left[\frac{d}{dt_A} H_R, H_3\right] = -\frac{i\hbar}{2} \left\{g''(T_B),H_R\right\} P_I,
\end{equation}
where $\{\cdot,\cdot\}$ denotes the anticommutator. This means that the dynamics of system $R$ is affected by the measurement of its energy by an internal system. Despite this, by using the Heisenberg equation of motion for the position $Q_I$ of the pointer, it holds that
\begin{equation}
    Q_I(t_f^A) - Q_I(t_i^A) = \int_{t_i^A}^{t_f^A} \left[\frac{1}{2} \left\{g(T_B(t_A)),H_B\right\} + g(T_B(t_A)) H_S + g(T_B(t_A))H_{int}(T_B(t_A))\right] dt_A,
\end{equation}
where the measurement was assumed to start and end at $t_i^A$ and $t_f^A$, respectively, depending on the uncertainty of $T_B$ from the perspective of $A$. Then, the shift in the pointer is proportional to the weighted average energy of system $R$ during the measurement's interaction had it not affected the system's dynamics. To be more precise, the previous result can be rewritten as
\begin{equation}
    Q_I(t_f^A) - Q_I(t_i^A) = K [H_B + H_S + \bar{H}_{int}],
\end{equation}
where
\begin{equation}
    \bar{H}_{int} \equiv \frac{1}{K} \int_{t_i^A}^{t_f^A} g(T_B(t_A)) H_{int}(T_B(t_A)) \, dt_A.
\end{equation}

Before concluding, observe that the disturbance to the evolution of system $R$ depends on derivatives of the measurement control function $g$. Most importantly, $g$ can be chosen in such a way that the net disturbance is null.

\end{document}